\documentclass[paper]{JHEP3} 

\usepackage{epsfig,multicol}

\newcommand\fverb{\setbox\pippobox=\hbox\bgroup\verb}
\newcommand\fverbdo{\egroup\medskip\noindent%
                        \fbox{\unhbox\pippobox}\ }
\newcommand\fverbit{\egroup\item[\fbox{\unhbox\pippobox}]}
\newbox\pippobox

\def\a{\alpha}
\def\b{\beta}
\def\c{\gamma}
\def\d{\delta}
\def\e{\epsilon}

\def\k{\kappa}
\def\l{\lambda}
\def\m{\mu}
\def\n{\nu}

\def\r{\rho}
\def\s{\sigma}

\def\D{\Delta}
\def\L{\Lambda}

\def\pl{\partial}

\def\Dslash{\,{\raise.15ex\hbox{/}\mkern-12mu D}}

\title{The Polarised Photon $g_1^\c$ Sum Rule at the Linear Collider and
High Luminosity B Factories \thanks{This research is supported in part by 
PPARC grant PP/G/O/2002/00470.  }}

\author{G.M. Shore\\
        Department of Physics\\
        University of Wales, Swansea\\
        Swansea SA2 8PP, U.K.\\
        E-mail: \email{g.m.shore@swansea.ac.uk}}

\preprint{SWAT 04-417}
 
\abstract{The sum rule for the first moment of the polarised (virtual) photon 
structure function $g_1^\c(x,Q^2;K^2)$ is revisited in the light of proposals 
for future $e^+ e^-$ colliders.
The sum rule exhibits an array of phenomena characteristic of QCD: for real photons
($K^2=0$) electromagnetic gauge invariance constrains the first moment to vanish;
the limit for asymptotic photon virtuality ($m_\r^2 \ll K^2 \ll Q^2$)
is governed by the electromagnetic $U_A(1)$ axial anomaly and the approach to
asymptopia by the gluonic anomaly; for intermediate values of $K^2$, it reflects 
the realisation of chiral symmetry and is determined by the off-shell radiative
couplings of the pseudoscalar mesons; finally, like many polarisation phenomena
in QCD, the first moment of $g_1^\c$ involves the gluon topological susceptibility.
In this paper, we review the original sum rule proposed by Narison, 
Shore and Veneziano and extend the relation with pseudoscalar mesons.
The possibility of measuring the sum rule in future polarised $e^+ e^-$ colliders
is then considered in detail, focusing on the International Linear Collider (ILC)
and high luminosity $B$ factories. We conclude that all the above features of the sum
rule should be accessible at a polarised collider with the characteristics of
SuperKEKB.
}

\begin{document}

\section{Introduction}

The sum rule for the first moment of the polarised photon structure function
$g_1^\c(x,Q^2;K^2)$ provides a window into many features of QCD dynamics,
including the gluonic axial anomaly and the realisation of chiral symmetry.
This sum rule was first proposed by Narison, Shore and Veneziano in 1992
\cite{NSVone} as part of a series of investigations into gluonic and 
anomaly-dependent phenomena in QCD, notably the origin of the `spin of the 
proton' suppression observed in the first moment of the polarised 
proton structure function $g_1^p$ 
\cite{SVpone,SVptwo,NSVpthree,SVpfour,DSVpfive,NSVpsix}.
At that time, however, the details of the sum rule were out of reach of 
contemporary colliders since the spin asymmetries which need to be measured 
to determine $g_1^\c(x,Q^2;K^2)$ require exceedingly large luminosities. 

Since that time, collider technology has moved on and plans are now well
advanced for machines capable of integrated annual luminosities in the
regime of inverse attobarns. It is therefore appropriate to revisit 
the $g_1^\c$ sum rule and investigate whether this new generation of colliders 
will be able to measure the full array of QCD phenomena encoded in it.

We focus on two future machines. First, the International Linear Collider (ILC)
has recently passed the technology choice phase and agreement has been found
to proceed with the `cold', i.e.~superconducting magnet, design. If international
agreement is forthcoming, it is hoped that a linear $e^+ e^-$ collider with
a CM energy of at least 500GeV will be commissioned around 2015. The projected
luminosity for the ILC is around $10^{34} {\rm cm}^{-2}{\rm s}^{-1}$,
corresponding to an annual integrated luminosity of order $0.1 {\rm ab}^{-1}$
\cite{ILCone,ILCtwo}.
Of course, for our purposes, the collider would have to be run in polarised mode.

The second machine we consider in detail is SuperKEKB. It was already noted
in ref.\cite{NSVone} that high-luminosity B factories were the colliders of
choice for measuring the $g_1^\c$ sum rule, since high energy is not in 
itself an advantage but ultra-high luminosity is essential. The proposed
upgrade of KEKB to SuperKEKB \cite{SuperKEKB}
envisages an 8GeV ($e^-$) on 3.5GeV ($e^+$)
collider with a target luminosity $5 \times 10^{35} {\rm cm}^{-2}{\rm s}^{-1}$,
corresponding to $5 {\rm ab}^{-1}$ annual integrated luminosity. As we shall
show, if this machine were run with polarised beams, this luminosity would allow
the full details of the off-shell first moment $g_1^\c$ sum rule to be
measured.

We come to these experimental considerations in section 5. We begin though
with a brief review of the derivation of the first moment sum rule itself,
emphasising that, as appropriate for a measurement in $e^+ e^-$ collisions 
(as opposed to doing two-photon physics using real back-scattered laser photons 
as the target), we are determining the {\it off-shell} structure function 
$g_1^\c(x,Q^2;K^2)$, where $K^2$ is the virtuality of the off-shell
`target' photon in DIS. Almost all the interesting QCD physics resides in the
$K^2$ dependence of the sum rule. An important point is that we should therefore
avoid an unnecessary use of the `equivalent photon' formalism \cite{EPone,EPtwo}
in establishing the sum rule. Moreover, all our results will be formulated 
in QCD field-theoretic terms, focusing on the OPE and current correlation 
functions, rather than using parton language. This makes the important 
non-perturbative results far more transparent.
(For a selection of reviews and recent papers on $g_1^\c$ from a parton
perspective, see 
e.g.~\cite{EPtwo,Partone,Parttwo,Partthree,Partfour,Partfive,Partsix,Partseven}.) 

Having reviewed the basic sum rule, in section 3 we study its `asymptotic' 
properties for momenta $K^2 \simeq 0$ and $K^2 \gg m_\r^2$. The first moment
is known \cite{Bassone,NSVone,Basstwo} 
to vanish for real photons as a consequence of
electromagnetic gauge invariance (current conservation). For photon
virtualities well above the relevant hadronic scale of $m_\r^2$ (but
still of course in the DIS regime $K^2 \ll Q^2$), $\int dx~g_1^\c(x,Q^2;K^2)$
tends to a value fixed by the electromagnetic axial $U_A(1)$ anomaly.
Moreover, the approach to this asymptotic region is governed by the
gluonic contribution to the anomaly, so there is much of theoretical 
interest even in this essentially perturbative regime. 

For intermediate virtualities, $K^2 \sim {\rm O}(m_\r^2)$, the first moment 
depends on the explicit momentum dependence of the form factors specifying
the three-current `AVV' correlation function involving the hadronic axial 
$U_A(1)$ current and two electromagnetic currents. This is an important 
non-perturbative object in QCD and the ability to measure it explicitly 
for a range of photon momenta would provide an interesting window into 
chiral symmetry breaking and associated QCD phenomena. The reasons for this 
are explored in considerable detail in ref.\cite{SVtwo}, a companion paper 
to ref.\cite{NSVone}. In particular, we can show that these form factors
are essentially the off-shell couplings of the pseudoscalar mesons
$\pi, \eta, \eta'$ to photons, whose on-shell limits are determined by the
radiative decays $\pi,\eta, \eta' \rightarrow \c \c$. The radiative pion
decay has played a distinguished role in establishing the reality of
anomalies and the nature of QCD (in particular, by providing a direct measure
of the number of colours). In the flavour singlet sector, the theory is even 
more interesting as it involves in an essential way the gluonic axial
$U_A(1)$ anomaly and the gluon content of the $\eta'$ meson \cite{SVeta}. 
In section 4, therefore, we explore the connection between the $g_1^\c$ sum rule
and radiative pseudoscalar decays, extending the results of ref.\cite{NSVone}
to incorporate the analysis developed in our 
papers \cite{Seta,Setamontp,Suppsala}.
One theoretically interesting feature is the link with the gluon topological 
susceptibility, which plays a key role in many polarised QCD phenomena,
notably the `spin of the proton'. This section may be read in conjunction
with another paper, ref.\cite{Snew}, in which we revisit our results 
\cite{Seta,Setamontp,Suppsala} for radiative pseudoscalar decays and 
their relation with the topological susceptibility and Witten-Veneziano 
formula \cite{W,V}, and derive explicit experimental
values for the pseudoscalar decay constants which may be (carefully) compared 
with large $N_c$ chiral Lagrangians \cite{Barc,KL} (see also ref.\cite{Feta}).

Having reviewed and developed the theory of the $g_1^\c$ first moment sum 
rule, we then return in section 5 to the experimental question of whether it
can be measured, including the full range of $K^2$ dependence, in the
forthcoming generation of high-luminosity colliders. Our conclusion is 
that the ILC is marginal for this purpose, but that a polarised collider
with the energy and luminosity of SuperKEKB would be able to uncover the 
full dynamical richness of the sum rule.

\section{The sum rule for $\int dx~g_1^\c(x,Q^2;K^2)$}

We are concerned with the process $e^+ e^- \rightarrow e^+ e^- X$, which at
sufficiently high energy is dominated by the two-photon interaction shown
in Fig.1. The deep-inelastic limit is characterised by $Q^2,\n_e,\n \rightarrow
\infty$ with\footnote{We have made a number of changes of notation compared
to ref.\cite{NSVone}. The dictionary is $K^2 \leftrightarrow \k^2$,
$\n_e \leftrightarrow \n$, $\n \leftrightarrow \tilde \n$, $x_e \leftrightarrow x$,
$x \leftrightarrow y$. The standard DIS notation ($\n,x$) therefore refers here
to the target photon, rather than the target electron as in ref.\cite{NSVone}}
$x_e = Q^2/2\n_e$ and $x= Q^2/2\n$ fixed, where (see Fig.1 for
definitions of the momenta) $Q^2 = -q^2,~ K^2 = -k^2,~\n_e = p_2.q,~ \n = k.q$
and $s=(p_1+p_2)^2$.
We also consider the `target photon' to be relatively soft, $K^2 \ll Q^2$.

\FIGURE
{\epsfxsize=6cm\epsfbox{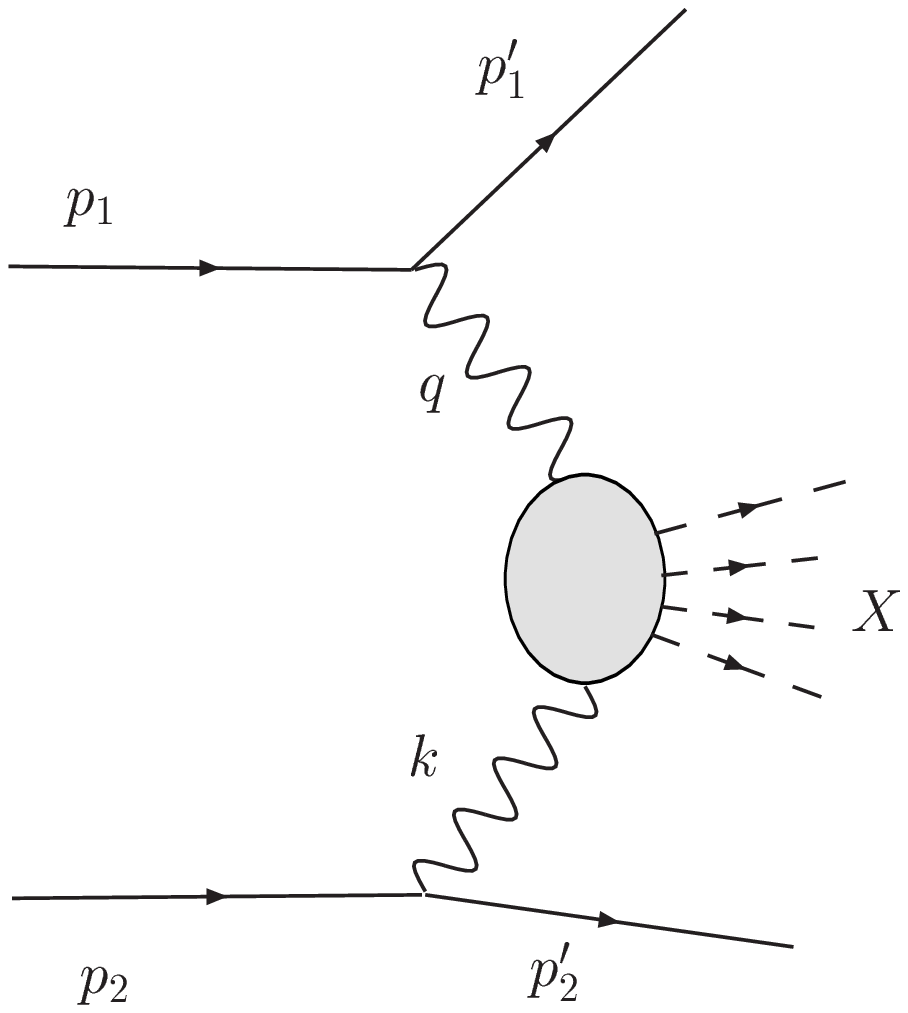} 
\caption{Kinematics for the two-photon DIS process 
$e^+ e^- \rightarrow e^+ e^- X$.}
\label{Fig:1}} 

Verifying the first moment sum rule for the polarised structure function
$g_1^\c(x,Q^2;K^2)$ requires studying
the spin asymmetries in cross-sections which are differential with respect to
$Q^2, K^2$ and $x_e$. Experimentally, these are determined from
\begin{eqnarray}
&\nonumber\\
x_e = {E_1' \sin^2{\theta_1\over2} \over E - E_1'\cos^2{\theta_1\over2}}~~~~~~
x = {Q^2\over Q^2+W^2} \nonumber\\
Q^2 = 4 E E_1' \sin^2{\theta_1\over2}~~~~~~~~
K^2 \simeq E E_2'\theta_2^2~~~~
\label{eq:ba}
\end{eqnarray}
Here, $E_1' (E_2')$ and $\theta_1' (\theta_2')$ are the energy and scattering angle 
of the hard-scattered (target) electron and $W$ is the invariant hadronic mass.
For the values $K^2\sim m_\r^2$ of interest in the sum rule, the target electron
is nearly-forward and $\theta_2'$ is very small. If it can be tagged, then the
virtuality $K^2$ is simply determined from eq.(\ref{eq:ba}); otherwise $K^2$ 
can be inferred indirectly from a measurement of the total hadronic energy.

A systematic presentation of the relations between cross-section moments
and structure functions from first principles may be found in ref.\cite{NSVone},
so here we shall only display some key formulae. `Electron structure functions'
$F_2^e(x_e,Q^2), ~F_L(x_e,Q^2)$ and $g_1^e(x_e,Q^2)$ are introduced in the
analogous way to ordinary nucleon structure functions and are related to
the spin-dependent cross-sections as follows:
\begin{equation}
\s~~=~~2\pi\a^2~{1\over s}~\int_0^\infty {dQ^2\over Q^2}\int_0^1
{dx_e\over x_e^2}~
\biggl[F_2^e \biggl({x_e s\over Q^2} -1 +{1\over2}{Q^2\over x_e s}\biggr)
- F_L^e {1\over2}{Q^2\over x_e s} \biggr] 
\label{eq:bb}
\end{equation}
\begin{equation}
\D\s~~=~~2\pi\a^2~{1\over s}~\int_0^\infty {dQ^2\over Q^2} \int_0^1 
{dx_e\over x_e}~
g_1^e \biggl[1- {1\over2}{Q^2\over x_e s}\biggr]
\label{eq:bc}
\end{equation} 
where $\s = {1\over 2}(\s_{++} + \s_{+-})$ and $\D\s = {1\over 2}(\s_{++} - 
\s_{+-})$ with $+,-$ referring to the electron helicities. The parameter
$Q^2/x_e s \ll 1$ and only leading order terms are 
retained below. $\a$ is the fine structure constant.

The photon structure functions themselves may be defined in the standard way
in terms of the matrix elements of the off-shell matrix elements
$\langle \c(k,\e^*)|J_\m^{\rm em}(q) J_\n^{\rm em}(-q)|\c(k,\e)\rangle$
of electromagnetic currents in the DIS limit. We can readily show that
the electron structure functions introduced above can be expressed as
convolutions of the photon structure functions with appropriate
Altarelli-Parisi splitting functions. In particular, we have
\begin{equation}
F_2^e(x_e,Q^2)~~=~~{\a\over2\pi}\int_0^\infty{dK^2\over K^2} \int_{x_e}^1
{dx\over x}{x_e\over x} P_{\c e}\Bigl({x_e\over x}\Bigr) ~F_2^\c(x,Q^2;K^2) 
\label{eq:bd} 
\end{equation}
\begin{equation}
g_1^e(x_e,Q^2)~~=~~{\a\over2\pi}\int_0^\infty{dK^2\over K^2} \int_{x_e}^1
{dx\over x} \D P_{\c e}\Bigl({x_e\over x}\Bigr) ~g_1^\c(x,Q^2;K^2) 
\label{eq:be}
\end{equation}
where 
\begin{equation}
P_{\c e}(z) = {1\over z} \bigl(1 + (1-z)^2\bigr) ~~~~~~~~~~~~~
\D P_{\c e}(z) = (2 - z)
\label{eq:bf}
\end{equation}

These results allow us to write relations that link the $x$-moments of 
the photon structure functions themselves to the $x_e$-moments of the 
cross-section. These key expressions, which we return to in section 5, are:
\begin{equation}
\int_0^1 dx_e ~x_e^n {d^3\s\over dQ^2 dx_e dK^2} ~~=~~
\a^3 {1\over Q^4 K^2}~\int_0^1 dz~z^n P_{\c e}(z)~
\int_0^1 dx~x^{n-1} F_2^\c(x,Q^2;K^2) 
\label{eq:bl}
\end{equation}
\begin{equation}
\int_0^1 dx_e ~x_e^n {d^3\D\s\over dQ^2 dx_e dK^2} ~~=~~
\a^3 {1\over s Q^2 K^2}~\int_0^1 dz~z^{n-1} \D P_{\c e}(z)~
\int_0^1 dx~x^{n-1} g_1^\c(x,Q^2;K^2) 
\label{eq:bm}
\end{equation}
The integrals factorise, so we see that the $n$th $x$-moments of the photon
structure functions are given by the $(n+1)$st $x_e$-moments of the fully
differential cross-sections.

The expressions ({\ref{eq:bd},\ref{eq:be}) are derived from first principles 
using only Feynman diagram rules and the operator product expansion. 
Despite the appearance of the AP splitting functions, no parton technology 
is used. Most importantly, it is not necessary to resort to the 
equivalent photon approximation \cite{EPone,EPtwo}
(see ref.\cite{NSVone} for a careful discussion of this point), so we can be
certain that our results are accurate throughout the required range of target 
photon momenta $K^2$. The relevant OPE is for the product of electromagnetic
currents:
$$
i~J_\m^{\rm em}(q) J_\n^{\rm em}(-q) ~~
\mathrel{\mathop{\sim}_{Q^2 \rightarrow\infty}}~~
\sum_{{n=1}{\rm ,odd}} \sum_a~ {2^n\over Q^{2n}} ~q_{\m_2} \ldots q_{\m_n}
i \e_{\m\n\a\m_1}q^\a ~E_1^{a,n}(Q^2) ~R_{a,n}^{\m_1 \ldots \m_n}(0) 
$$
\begin{equation}
{}~~~~~~~~~~~~~~~~~~~~~~~~~~~~~+~~~ {\rm even-parity}~{\rm operators}
\label{eq:bg}
\end{equation}
where the $E_1^{a,n}(Q^2)$ are Wilson coefficients and $R_{a,n}^{\m_1 
\ldots \m_n}(0)$ are the complete set of odd-parity, twist 2 operators in QCD.
(We only indicate here the odd-parity operators, which contribute to $g_1^\c$,
to establish notation; the even-parity operators contribute to $F_2^\c$
and $F_L^\c$. See ref.\cite{NSVone} for full details and the explicit 
forms of the relevant operators.)
Their form factors in the 3-point correlation functions with the photon
fields $A_\l(k)$ are:
\begin{equation}
\langle 0|R_{a,n}^{\m_1 \ldots \m_n}(0) A_\l(k) A_\r(-k)|0\rangle ~~=~~
{1\over K^4} k^{\m_2} \ldots k^{\m_n} i \e_{\l\r}{}^{\m_1\a} k_\a ~ 
\hat R_{a,n}(K^2)~~~~~~~~(n\ge 1, {\rm odd})
\label{eq:bh}
\end{equation}

The key result now is the relation between the moments of the structure
functions and the Wilson coefficients and form factors from the OPE.
We can show \cite{NSVone}
\begin{equation}
\int_0^1 dx~x^{n-1}~g_1^\c(x,Q^2;K^2) ~~=~~ \sum_a ~E_1^{a,n}(Q^2)~
\hat R_{a,n}(K^2)
\label{eq:bi}
\end{equation}
with similar expressions for $F_2^\c$ and $F_L^\c$. We are of course
primarily concerned with the $n=1$ moment of $g_1^\c(x,Q^2;K^2)$.
In this case, the only relevant $R_{a,1}^\m$ operator is the hadronic 
$U_A(1)$ axial current $J_{\m5}^r$,
\begin{equation}
R_{a,1}^\m ~\rightarrow ~J_{\m5}^r ~=~ \bar\psi T^r \c^\m \c_5 \psi
\label{eq:bj}
\end{equation}
where $T^r$ are the $SU(3)$ generators (including the flavour singlet 
$T^0 = {\bf 1}$), and $\psi$ are quark fields. Clearly, only the diagonal
generators $a=3,8,0$ contribute. The form factor is therefore defined as
\begin{equation}
\langle 0|J_{\m 5}^r(0) A_\l(k) A_\r(-k)|0\rangle ~=~ {1\over K^4} i
\e_{\l\r\m\a} k^\a ~ \hat R_{r,1}(K^2)
\label{eq:bk}
\end{equation}
and the essential first moment sum rule is
\begin{equation}
\int_0^1 dx~g_1^\c(x,Q^2;K^2) ~~=~~ \sum_{r=3,8,0} ~E_1^{r,1}(Q^2)~
\hat R_{r,1}(K^2)
\label{eq:bn}
\end{equation}

To sum up this section, we have established in eq.(\ref{eq:bm}) how to
measure the first moment of the polarised photon structure function 
$g_1^\c(x,Q^2;K^2)$ in terms of the spin asymmetry of the differential
cross-section, $d^3\D\s /dQ^2 dx_e dK^2$. The kinematic variables $Q^2$ and
$x_e$ are found from the energy and scattering angle of the hard-scattered
electron, while the target photon virtuality $K^2$ is most easily 
found by tagging the nearly-forward electron. On the theory side,
eq.(\ref{eq:bn}) determines the first moment of $g_1^\c(x,Q^2;K^2)$ in terms
of a perturbatively known Wilson coefficient $E_1^{r,1}(Q^2)$ and
a non-perturbative form factor $\hat R_{r,1}(K^2)$ characterising
the AVV 3-current correlation function involving the hadronic $U_A(1)$
axial current and two electromagnetic currents.

\section{The AVV correlation function, anomalies and momentum 
dependence}

The first element of the sum rule is the Wilson coefficient $E_1^{r,1}(Q^2)$.
The $Q^2$-dependence is governed by the RGE and the solution is well known.
For QCD with $N_c=3$ and $N_f=3$ active flavours (results for the arbitrary
$n$th moments and general $N_c,N_f$ are quoted in ref.\cite{NSVone}), we have
\begin{equation}
E_1^{r,1} ~=~ c^{(r)}~\Bigl(1- {\a_s(Q^2)\over\pi}\Bigr)~~~~~~~~~~~~~~~(r=3,8) 
\label{eq:caa}
\end{equation}
\begin{equation}
E_1^{0,1} ~=~ c^{(0)}~\exp\Bigl[\int_0^t dt' ~\c(\a_s(t'))\Bigr]~
\Bigl(1- {\a_s(Q^2)\over\pi}\Bigr)
\label{eq:ca}
\end{equation}
where $t={1\over2}\ln Q^2/\m^2$. The coefficients are fixed by the quark charges
and are $c^{(3)}= 1/3$, $c^{(8)}=1/3\sqrt3$ and $c^{(0)}=2/9$. Since the
singlet axial current is not conserved because of the gluonic contribution
to the $U_A(1)$ anomaly, it is associated with an anomalous dimension $\c$,
which enters into the momentum dependence of the singlet Wilson coefficient.
Explicitly, $\c = -\c_0{\a_s\over4\pi} - \c_1 {\a_s^2\over(4\pi)^2} + \ldots$
where $\c_0 = 0$ and $\c_1 = 3/4$. Notice the important result that the 
expansion begins only at $O(\a_s^2)$. The beta function, which determines the 
running of the QCD coupling $\a_s(Q^2)$, is similarly given by
$\b = -\b_0{\a_s^2\over4\pi} - \b_1 {\a_s^3\over(4\pi)^2} + \ldots$
with $\b_0 = 18$. 

Now consider the amputated AVV correlation function in eq.(\ref{eq:bk}), 
defined with the electromagnetic current $J_\l^{(\rm em)}$. We first 
allow the axial current momentum $p$ to be non-zero and subsequently take
the required limit $p\rightarrow 0$. There are two important Ward identities.
Electromagnetic current conservation implies
\begin{equation}
i k_1^\l \langle 0|J_{\m5}^r(p) J_\l^{(\rm em)}(k_1)  
J_\r^{(\rm em)}(k_2)|0\rangle ~~=~~0 ~~~~~~~~~~~
({\rm sim~ for}~ k_2^\r)
\label{eq:cb}
\end{equation}
The (anomalous) chiral Ward identity, which follows from the usual
anomalous conservation law (for $N_f=3$)
\begin{equation}
\pl^\m J_{\m5}^r ~=~ M_{rs}\phi_5^s + 6 Q \d_{r0} +  a^{(r)} {\a\over 8\pi}
\tilde F^{\m\n} F_{\m\n}
\label{eq:ccc}
\end{equation}
where $\phi_5^r = \bar\psi T^r \c_5\psi$ and $Q= {\a_s\over8\pi}{\rm tr}
\tilde G^{\m\n}G_{\m\n}$, with $G_{\m\n}$ the gluon field strength and  
$F_{\m\n}$ the electromagnetic field strength, is
$$
i p^\m \langle0| J_{\m5}^r(p)  J_\l^{(\rm em)}(k_1)  J_\r^{(\rm em)}(k_2)
|0\rangle
~-~ M_{rt}~\langle0| \phi_5^t(p)  J_\l^{(\rm em)}(k_1)  
J_\r^{(\rm em)}(k_2)|0\rangle 
$$
\begin{equation}
~~~~~~~~~~~~~~~~~~~-~\d^{r0}~ 6 ~\langle0| Q(p)  J_\l^{(\rm em)}(k_1)  
J_\r^{(\rm em)}(k_2)|0\rangle ~+~{1\over 8\pi^2} a^{(r)} 
\e_{\l\r\a\b}k_1^\a k_2^\b ~~=~~0
\label{eq:cc}
\end{equation}
The notation used for the quark masses follows
ref.\cite{Seta}:~ 
${\rm diag}(m_u,m_d,m_s) = \sum_{r=3,8,0} m_r T^r$, then $M_{rt} = d_{rst}m_s$,
where $d_{rst}$ are the usual $SU(3)$ $d$-symbols. 
The term involving the gluon topological
density $Q$ is the gluonic $U_A(1)$ anomaly and only appears in the flavour
singlet case, while the final term arises because of the electromagnetic
$U_A(1)$ anomaly whose strength depends on the quark charges. With our
normalisations, $a^{(3)} = 1$, $a^{(8)}=1/\sqrt3$ and $a^{(0)}=4$.
(Notice that this notation differs from ref.\cite{NSVone}. $a^{(r)}$ here 
is $2N_c$ times the $a^{(r)}$ of ref.\cite{NSVone} but coincides
with the $a_{\rm em}^r$ of refs.\cite{Seta,Setamontp,Suppsala}.)

Now define form factors for the correlation functions appearing above:
\begin{eqnarray}
&\nonumber\\
&-i\langle0|J_{\m5}^r(p) J_\l^{(\rm em)}(k_1)  J_\r^{(\rm em)}(k_2)|0\rangle ~=~ 
A_1^r ~\e_{\m\l\r\a}k_1^\a ~+~ A_2^r ~\e_{\m\l\r\a}k_2^\a ~~~~~~~~~~~~~
\nonumber\\
&~~~~~~~~~~~~~~~~~~~~~~~~~~~~~~~~~~~~~~~~~
+A_3^r ~\e_{\m\l\a\b}k_1^\a k_2^\b k_{2\r} ~+~ 
A_4^r ~\e_{\m\r\a\b}k_1^\a k_2^\b k_{1\l} \nonumber\\
&~~~~~~~~~~~~~~~~~~~~~~~~~~~~~~~~~~~~~~~~~
+A_5^r ~\e_{\m\l\a\b}k_1^\a k_2^\b k_{1\r} ~+~
A_6^r ~\e_{\m\r\a\b}k_1^\a k_2^\b k_{2\l}
\label{eq:cd}
\end{eqnarray}
where the six form factors are functions of the invariant momenta, i.e.
$A_i^r = A_i^r(p^2,k_1^2,k_2^2)$,
\begin{equation}
M_{rt}~\langle 0|\phi_5^t(p) J_\l^{(\rm em)}(k_1) 
J_\r^{(\rm em)}(k_2)|0\rangle ~=~ 
D^r(p^2,k_1^2,k_2^2) ~\e_{\l\r\a\b} k_1^\a k_2^\b
\label{eq:ce}
\end{equation}
\begin{equation}
6~\langle 0|Q(p)  J_\l^{(\rm em)}(k_1)  J_\r^{(\rm em)}(k_2)|0\rangle ~=~ 
B(p^2,k_1^2,k_2^2) ~\e_{\l\r\a\b} k_1^\a k_2^\b
\label{eq:cf}
\end{equation}
With these definitions, it follows immediately from eq.(\ref{eq:bk}) that
the second element of the sum rule, i.e.~the form factor $\hat R_{r,1}$, is just
\begin{eqnarray}
&\nonumber\\
&\hat R_{r,1}(K^2) ~~=~~ 4\pi\a~ \bigl(A_1^r - A_2^r\bigr)(K^2)~~~~~~~~~~~~~~
\nonumber\\
&~~~~~~~~~~~~~~~~~~~~~~~~=~~- 4\pi\a~ \Bigl( D^r(K^2) + \d^{r0} B(K^2) 
-{1\over8\pi^2}a^{(r)}
\Bigr)
\label{eq:cg}
\end{eqnarray}
where we have used the notation $D^r(K^2) = D^r(0,k^2,k^2)$ etc.
The non-perturbative QCD dynamics governing the first moment sum rule is 
therefore encoded in these 3-current form factors.

In the next section, we discuss the $K^2$ dependence of the sum rule in terms of
these form factors. However, without any detailed knowledge of their 
non-perturbative features, we can already determine the first moment of 
$g_1^\c$ in the limit $K^2=0$, corresponding to real photons, and in 
the `asymptotic' region $K^2 \gg m_\r^2$.

The first observation is that electromagnetic current conservation requires
$\hat R_{r,1}(K^2)$ to vanish at $K^2=0$. Substituting the form factor expansion
(\ref{eq:cd}) into the Ward identity (\ref{eq:cb}), and taking $p=0$, we find
\begin{eqnarray}
&\nonumber\\
&A_1^r ~~=~~ A_3^r k_2^2 + A_5^r {1\over2}\bigl(p^2 - k_1^2 - k_2^2\bigr) 
\nonumber\\
&A_2^r ~~=~~ A_4^r k_1^2 + A_6^r {1\over2}\bigl(p^2 - k_1^2 - k_2^2\bigr)
\label{eq:ch}
\end{eqnarray}
Provided none of the form factors have singularities at $p^2=0$ (or $K^2=0$),
as is the case away from the chiral limit, then it follows immediately
that both $A_1^r(K^2)$ and $A_2^r(K^2)$ are of $O(K^2)$ for small photon 
virtuality. (The chiral limit is subtle and is discussed in detail in 
ref.\cite{SVtwo}.) We therefore establish $\hat R_{r,1}(0) = 0$ and therefore
\cite{Bassone,NSVone,Basstwo}
\begin{equation}
\int_0^1 dx~g_1^\c(x,Q^2;K^2=0) ~~=~~ 0
\label{eq:ci}
\end{equation}

Next, we consider the asymptotic limit of large $K^2$, while still keeping in 
the DIS regime of $K^2 \ll Q^2$. For this, we need the large $K^2$ limit of the
form factors $A_i^r(K^2)$, $D^r(K^2$ and $B(K^2)$, which can be obtained using 
the renormalisation group. The flavour non-singlet ($r=3,8)$ and singlet ($r=0$)
cases are different. In the non-singlet case, since the axial current is
conserved, the form factors $A_i^r(K^2)$ satisfy a homogeneous RGE 
(see ref.\cite{NSVone} for explicit details), with the standard solution
\begin{equation}
A_i^r(K^2; \a_s(\m); m)~~=~~A_i^r(\m^2; \a_s(t); e^{-t}m(t)~)
\label{eq:cj}
\end{equation}
where $\m$ is an RG reference scale, $t = {1\over2}\ln{K^2\over\m^2}$ (contrast
with the `$t$' in the Wilson coefficient expressions, which refers to the scale 
$Q^2$), $\a_s(t)$ and $m(t)$ are running couplings and $m$ generically 
denotes the individual quark masses $m^r$. The large $K^2$ limit of
$A_1^r - A_2^r$ is therefore obtained from the correlation function  
evaluated at weak coupling in the chiral limit. In this limit, $D^r$ is clearly
zero, so recalling eq.(\ref{eq:cg}), we conclude that in the
flavour non-singlet sector,
\begin{equation}
\hat R_{r,1}(K^2\rightarrow\infty) ~~=~~ {1\over2}  a^{(r)}~{\a\over\pi} 
~~~~~~~~~~~~~(r=3,8)
\label{eq:ck}
\end{equation}
The asymptotic value of the form factor is therefore determined by the 
electromagnetic $U_A(1)$ anomaly coeffiecient.

In the flavour singlet case, however, the 3-current correlation function
satisfies an inhomogeneous RGE with anomalous dimension $\c$ because of the 
anomalous non-conservation of the singlet axial current. In this case, therefore,
\begin{equation}
A_i^0(K^2;\a_s(\m); m) ~~=~~ \exp\Bigl[-\int_0^t dt'~\c(\a_s(t'))\Bigr]~
A_i^0(\m^2; \a_s(t); e^{-t}m(t))
\label{eq:cl}
\end{equation}
Here, we need both the form factors $D^0$ and $F^0$ 
to evaluate the r.h.s.
At weak coupling, the correlation function involving the topological charge $Q$
is of $O(\a_s^2)$, and so contributes only at the same order as other neglected
terms. So once again, the asymptotic limit is controlled simply by the
anomaly coefficient. However, this time we also need the anomalous dimension
term, and the final result is
\begin{equation}
\hat R_{0,1}(K^2\rightarrow\infty) ~~=~~ {1\over2} a^{(0)}~{\a\over\pi}~
\exp\Bigl[-\int_0^t dt'~\c(\a_s(t'))\Bigr]
\label{eq:cm}
\end{equation}

The asymptotic form for the $g_1^\c$ sum rule is finally obtained by putting 
together eqs.(\ref{eq:ck}),(\ref{eq:cm}) for the form factors with
eqs.(\ref{eq:caa}),(\ref{eq:ca}) for the Wilson coefficients.
This gives:
\begin{equation}
\int_0^1 dx~g_1^\c(x,Q^2;K^2\rightarrow \infty) ~~=~~ \sum_{r=3,8,0} ~
E_1^{r,1}(Q^2)~\hat R_{r,1}(K^2\rightarrow \infty)~~~~~~~~~~~~~~ 
\end{equation}
\begin{equation}
~~~~~~~~~~~~~~~~~~~~~~~~~~~=~~
{1\over2} {\a\over\pi} ~\Bigl(1 - {\a_s(Q^2)\over\pi}\Bigr)~\biggl[
c^{(3)} a^{(3)} + c^{(8)} a^{(8)} + c^{(0)}a^{(0)}
\exp\Bigl[\int_{t(K)}^{t(Q)} dt'~\c(\a_s(t'))\Bigr]~ \biggr]
\label{eq:cn}
\end{equation}
with the obvious notation $t(Q) = {1\over2} \ln{Q^2\over\m^2}$,
$t(K) = {1\over2} \ln{K^2\over\m^2}$. 
Substituting ${\a_s(t)\over4\pi} \simeq {1\over \b_0 t}$ for the running 
couplings and reorganising terms, we obtain the final form of the sum rule:
\begin{equation}
\int_0^1 dx~g_1^\c(x,Q^2;K^2\rightarrow \infty) ~~=~~ {2\over3}{\a\over\pi} 
\biggl[1 ~-~ {4\over9}{1\over \ln Q^2/\L^2} ~+~
{16\over81} \biggl({1\over \ln Q^2/\L^2}- {1\over \ln K^2/\L^2}
\biggr) \biggr]
\label{eq:co}
\end{equation}
Notice that the overall normalisation factor is $N_c \sum_f \hat e_f^4$,
proportional to the sum of the fourth power of the quark charges $\hat e_f$,
corresponding to the lowest order box diagram contributing to $g_1^\c$.

The key physics in eqs.(\ref{eq:cn}),(\ref{eq:co}) is that the
first moment of $g_1^\c(x,Q^2;K^2)$ in the asymptotic limit $m_\r^2\ll K^2\ll 
Q^2$ for the target photon virtuality is governed by the quark charges, with a
flavour dependence reflecting the electromagnetic $U_A(1)$ anomaly coefficients.
The approach to this asymptotic value depends on logarithmic corrections
given by the anomalous dimension arising from the gluonic $U_A(1)$
anomaly in the flavour singlet current.

In between the limits $K^2=0$ and $K^2 \gg m_\r^2$, the sum rule depends on
form factors $F^r(K^2)$ as follows (substituting for $c^{(r)}$ and $a^{(r)}$ 
compared to eq.(\ref{eq:cn})):
$$
\int_0^1 dx~g_1^\c(x,Q^2;K^2) ~~=~~
{1\over18}{\a\over\pi} ~\Bigl(1 - {\a_s(Q^2)\over\pi}\Bigr)~~~~~~~~~~~~~~~~~~~~~
~~~~~~~~~~~~~~~~~~~~~~~~~~
$$
\begin{equation}
~~~~\times~\biggl[
3 F^3(K^2) +  F^8(K^2) + 8 F^0(K^2;\m^2=K^2)
\exp\Bigl[\int_{t(K)}^{t(Q)} dt'~\c(\a_s(t'))\Bigr]~ \biggr]
\label{eq:cp}
\end{equation}
The form factors $F^r(K^2)$ interpolate between 0 for $K^2=0$ and
1 for asymptotically large $K^2$. Notice that in the anomalous singlet 
sector, we have to specify the renormalisation scale -- with the
anomalous dimension factor as shown, the form factor $F^0(K^2)$ must
be evaluated at $\m^2=K^2$.

These form factors are simply written in terms of those defined in 
eqs.(\ref{eq:cd}),(\ref{eq:ce}),(\ref{eq:cf}). Explicitly,
$$
F^r(K^2) ~~=~~\Bigl({1\over8\pi^2}a^{(r)}\Bigr)^{-1}~
(A_1^r - A_2^r)(K^2)~~~~~~~~~~~~
$$
\begin{equation}
~~~~~~~~~~~~~~~~=~~1- \Bigl({1\over8\pi^2}a^{(r)}\Bigr)^{-1}~
\Bigl(D^r(K^2) + \d^{r0}B(K^2)\Bigr)
\label{eq:cq}
\end{equation}
The full $K^2$ dependence of the sum rule is therefore governed by the
AVV correlation function, or alternatively, the correlation functions
in the corresponding Ward identity.
The sum rule therefore gives an experimental measure of these correlators,
which are sensitive to the realisation of chiral symmetry in QCD \cite{SVtwo}
and the gluonic $U_A(1)$ anomaly. A first principles calculation of these
correlation functions in QCD, if it were possible, would therefore give a 
complete prediction for the first moment sum rule.

\section{$U_A(1)$ PCAC, ~$\pi,\eta,\eta'\rightarrow \c\c$,  
and the gluon topological susceptibility}

In order to gain some more insight into the non-perturbative behaviour of
the first moment of $g_1^\c(x,Q^2;K^2)$, we can use the ideas of PCAC
and spontaneously broken chiral symmetry to rewrite the sum rule in terms 
of the off-shell couplings for radiative decays of the pseudo-Goldstone
bosons $\pi, \eta$ and $\eta'$, since these are also controlled by the 
AVV correlation function in QCD. Of course, the gluonic anomaly makes
the application of PCAC to the $U_A(1)$ sector both interesting and
subtle. In particular, the sum rule is sensitive
to the gluon topological susceptibility, which plays a key role in
many polarisation-dependent phenomena in QCD.

The link between the AVV correlation function and radiative pseudoscalar decays
arises by writing the close analogue of the axial Ward identity (\ref{eq:cc})
involving photon states:
\begin{equation}
ip^\m \langle 0|J_{\m5}^r|\c\c\rangle ~~=~~
M_{rt} \langle 0| \phi_5^t|\c\c\rangle ~+~
6 \d_{r0}\langle 0|Q|\c\c\rangle ~+~ 
 a^{(r)} {\a\over8\pi}\langle 0|F^{\m\n}\tilde F_{\m\n}|\c\c\rangle
\label{eq:daa}
\end{equation}
and assuming pseudoscalar dominance of the matrix elements to rewrite
them in terms of the radiative couplings $g_{\pi\c\c}$, $g_{\eta\c\c}$
and $g_{\eta'\c\c}$. However, because of the anomaly, 
the relation of the operators $\phi_5^r$ and $Q$
to the physical pseudoscalars $\pi,\eta,\eta'$ is not entirely straightforward
and it is best to make a change of variables to operators which are more
appropriate as interpolating fields for the pseudoscalar particles.
This approach to `$U_A(1)$ PCAC' and radiative pseudoscalar decays 
is described in detail in refs.\cite{Seta,Setamontp,Suppsala}. See also 
ref.\cite{Snew} for an analysis of experimental values for the various
decay constants which arise.

The result is the following set of expressions for
the form factors $F^r(K^2)$ in terms of the {\it off-shell} radiative
pseudoscalar couplings for photon virtuality $K^2$:
$$
F^3(K^2) ~~=~~ 1 - \Bigl( {\a\over\pi}\Bigr)^{-1}~f_\pi g_{\pi\c\c}(K^2)
$$
$$
F^8(K^2) ~~=~~ 1 - \Bigl({1\over\sqrt3} {\a\over\pi}\Bigr)^{-1}~
\Bigl(f^{8\eta} g_{\eta\c\c}(K^2) + f^{8\eta'}g_{\eta'\c\c}(K^2) \Bigr)
$$
\begin{equation}
F^0(K^2;\m^2) ~~=~~ 1 - \Bigl(4 {\a\over\pi}\Bigr)^{-1}~
\Bigl(f^{0\eta}g_{\eta\c\c}(K^2) + f^{0\eta'} g_{\eta'\c\c}(K^2) 
+ 6A g_{G\c\c}(K^2;\m^2) \Bigr)
\label{eq:dk}
\end{equation}

We now discuss what insight this new representation gives into the
momentum dependence of the form factors $F^r(K^2)$. The first striking
observation is that the first moment of $g_1^\c(x,Q^2;K^2)$ for an
off-shell photon target involves the gluon topological susceptibility,
as is characteristic of many polarisation phenomena in QCD. This arises
through the dependence of the singlet form factor $F^0(K^2)$ 
on the non-perturbative constant $A$ which controls the topological 
susceptibility. For non-vanishing quark masses \cite{Top}: 
\begin{equation}
\chi(0) ~~\equiv ~~ \langle Q Q\rangle ~~=~~
-A \biggl(1-A\sum_q{1\over m_q \langle\bar q q\rangle}\biggr)^{-1}
\label{eq:dkk}
\end{equation}
The corresponding radiative coupling $g_{G\c\c}$ has a clear theoretical 
interpretation as the coupling of the two-photon state to a glueball-like 
operator $G$ orthogonal to the physical $\eta'$. It does not, however,
necessarily refer to a physical particle state (see 
e.g.~refs.\cite{Seta,Setamontp,Suppsala,Snew} for a further discussion), 
so we do not have a clear intuition about its momentum dependence.

Although these anomalous contributions are interesting from a theoretical
perspective, in practice they may not be so significant for the sum rule.
Arguments based on the $1/N_c$ expansion or OZI rule, carefully applied
to the flavour singlet channel, suggest that the contribution
of the $6Ag_{G\c\c}$ term on the l.h.s.~of eq.(\ref{eq:dk}) is subdominant.
An explicit fit \cite{Snew} of the decay constants and couplings in 
eqs.(\ref{eq:dk}) indeed confirms that, for on-shell photons, the relative 
contribution of this term is around $20\%$. 

The main result implied by eqs.(\ref{eq:dk}) is that the
momentum dependence of the form factors in the sum rule is determined
by the non-perturbative couplings $g_{\pi\c\c}(K^2)$, $g_{\eta\c\c}(K^2)$ and
$g_{\eta' \c\c}(K^2)$. The relevant mass scale determining the crossover 
from $F^r(0) = 0$ to $F^r(\infty) = 1$ is therefore given by the
non-perturbative scale in the photon channel of the pseudoscalar 
radiative coupling. Well-established ideas invoking vector meson dominance
(VMD) imply that for $F^3(K^2)$ this scale is $m_\r^2$ (equivalently 
$m_\phi^2$, $m_\omega^2$ for the other flavours). Given that QCD
spontaneously breaks chiral symmetry, we therefore expect that the 
form factors interpolate smoothly between 0 and 1 with a crossover
scale characterised by the vector meson masses. This is in sharp 
contrast to a perturbative QCD picture, in which this scale would
correspond instead to the light quark masses \cite{CCM}. 
(See ref.\cite{SVtwo} for an extensive discussion of this point.\footnote{
Of course, for final states characteristic of heavy quarks (c,b) with 
mass $> \L_{\rm QCD}$, the crossover scale would simply be the quark
mass itself, $m_{\rm c,b}^2$. Similarly for leptonic final states
which probe the QED structure of the photon.}) 
The $K^2$ dependence
of the first moment sum rule for $g_1^\c(x,Q^2;K^2)$ is therefore
a clear signal of chiral symmetry breaking. 

We can try to justify this VMD prediction directly from QCD field theory 
as follows. (See also the closely related analyses of the AVV correlation
functions and radiative and leptonic pseudoscalar decays in 
refs.\cite{SNPaver,Knecht,Pich} (and references therein). Comprehensive reviews of
relevant QCD sum rule results may be found in refs.\cite{SNone,SNtwo}.) 
Once we have related the form factors to the pseudoscalar
radiative couplings, we can use the OPE for the two electromagnetic
currents in, for example, the matrix element $\langle \pi|J_\l^{(\rm em)}(k)
J_\r^{(\rm em)}(-k)|0\rangle$ for large $K^2$ (compare eq.(\ref{eq:bg}))
to write
\begin{equation}
\langle \pi|J_\l^{(\rm em)}(k) J_\r^{(\rm em)}(-k)|0\rangle ~~=~~
2 \e^\m{}_{\a\l\r} k^\a {1\over K^2} E_1^{3,1}(K^2) 
\langle \pi|J_{\m5}^3(0)|0\rangle
~~+~~\ldots
\label{eq:dl}
\end{equation}
Comparing with the definitions of the form factors and couplings, we find 
to leading order,
\begin{equation}
F^3(K^2 \rightarrow\infty) ~~=~~ 1 - {(4\pi)^2\over3} f_\pi^2 {1\over K^2} 
~~+~~\ldots
\label{eq:dm}
\end{equation}
Comparing this large $K^2$ behaviour with a simple interpolation formula 
such as $F^3(K^2) \sim K^2/(K^2+M^2)$, we would identify the characteristic
crossover mass scale as $M^2 \sim {(4\pi)^2\over3}f_\pi^2$, which is
numerically $\sim m_\r^2$. This estimate is therefore consistent
with the VMD picture.

This completes our discussion of the non-perturbative QCD dynamics
behind the momentum dependence of the $g_1^\c(x,Q^2;K^2)$ sum rule.
In the next section, we move on to discuss the experimental question of
whether all this can actually be directly measured in DIS experiments
at $e^+ e^-$ colliders.

\section{Cross-sections and spin asymmetries at the ILC and SuperKEKB}

The spin-dependent cross-sections for the two-photon DIS process 
$e^+ e^- \rightarrow e^+ e^- X$ defined in section 2 are given in 
ref.\cite{NSVone} as
\begin{equation}
\s ~~\simeq~~ 
{\a^4\over2\pi}~ \bar a~ {1\over Q^2_{\rm min}}\log{Q^2_{\rm min}\over\L^2}
\log{K^2_{\rm max}\over K^2_{\rm min}}
\log{x_e^{\rm max}\over x_e^{\rm min}}
\log{x^{\rm max}\over \langle x_e\rangle}
\label{eq:ea}
\end{equation}
and 
\begin{equation}
\D \s ~~\simeq~~
{\a^4\over2\pi}~ \bar b ~{1\over s} 
\log{Q^2_{\rm max}\over Q^2_{\rm min}}
\log{\langle Q^2\rangle\over\L^2}
\log{K^2_{\rm max}\over K^2_{\rm min}}
\log{x_e^{\rm max}\over x_e^{\rm min}}
\log{x^{\rm max}\over \langle x_e\rangle}
\label{eq:eb}
\end{equation}
In these expressions, we have included experimental cuts on the maximum and 
minimum values of the kinematical variables $Q^2,K^2,x_e$ and $x$.
$\langle Q^2\rangle$ is the geometric mean of $Q^2_{\rm max}$ and
$Q^2_{\rm min}$ (similarly for $\langle x_e\rangle$). The constants $\bar a$ and
$\bar b$ are approximations to the functions $a(x)$ and $b(x)$ given by
inverse Mellin transforms of the moments $a_n$ and $b_n$ corresponding to
the higher spin operators in the OPEs for $F_2^\c$ and $g_1^\c$. Numerically,
$\bar a \simeq \bar b \simeq 1.5$. 

The spin asymmetry is therefore
\begin{equation}
{\D\s\over\s}~~\simeq~~
{1\over2}{Q^2_{\rm min}\over s}
\log{Q^2_{\rm max}\over Q^2_{\rm min}}
\biggl[1 + \log{Q^2_{\rm max}\over\L^2}
\biggl(\log{Q^2_{\rm min}\over\L^2}\biggr)^{-1}\biggr]
\label{eq:ec}
\end{equation}
In order to extract information on the $g_1^\c$ structure function from $\D\s$,
we need this spin asymmetry to be large. More precisely, for a statistically
significant result, we require $\D\s/\s \ge 1/\sqrt{L\s}$, where $L$ is the 
integrated luminosity \cite{NSVone}.

Experimentally, the accelerator design specifies the CM energy $s$ and
luminosity $L$, but we can then choose the cuts on the kinematic variables, 
subject of course to detector constraints, in order to maximise the
measured cross sections and spin asymmetries necessary to determine $g_1^\c$.
The relevant cuts are on $Q^2, K^2, \n_e$ and $\n$. The upper cut on $Q^2$ is
limited by the detector acceptance and we take $Q^2_{\rm max} \simeq s/4$. For the 
lower cut, $Q^2_{\rm min}$, we have to be within the DIS region but otherwise will
keep this as a free parameter to be varied to try and obtain the most
statistically significant measurement. For $K^2$, we set a lower cut at
$K^2\simeq m_e^2$ and vary to an upper limit well above $m_\r^2$, taking
$K^2_{\rm max} = 1{\rm GeV}^2$ in the total cross-section estimates. 
Since $\n = {1\over2}(Q^2 + W^2)$, and $W^2_{\rm min}$
is small, we choose the following cuts on the Bjorken variables:
$\n_e^{\rm min} = \n^{\rm min} = {1\over2}Q^2_{\rm min}$ and
$\n_e^{\rm max} = \n^{\rm max} = {1\over2}s$. Inserting these cuts into the
formulae for the total cross-section and spin asymmetry, we have
\begin{equation}
\s ~~\simeq~~
0.5 \times 10^{-8}~{1\over Q^2_{\rm min}}~\log{Q^2_{\rm min}\over \L^2}
~\biggl(\log{s\over Q^2_{\rm min}}\biggr)^2
\label{eq:ed}
\end{equation}
and 
\begin{equation}
{\D\s\over\s} ~~=~~
{1\over2}~{Q^2_{\rm min}\over s}~ \log{s\over 4 Q^2_{\rm min}}~
\biggl[ 1 + \log{s\over 4\L^2}\biggl(\log{Q^2_{\rm min}\over\L^2}\biggr)^{-1}
\biggr]
\label{eq:ee}
\end{equation}

As noted in section 2, the $n$th $x$-moment of $g_1^\c$ is determined by 
the $(n+1)$st $x_e$-moment of $\D\s$. In particular, from eq.(\ref{eq:bm}) 
we have
\begin{equation}
\int_0^1 dx_e~x_e {d^3\D\s\over dQ^2 dx_e dK^2} ~~=~~
{3\over2} \a^3 {1\over s Q^2 K^2}
\int_0^1 dx~g_1^\c(x,Q^2;K^2)
\label{eq:ef}
\end{equation}
Comparing with the first moment sum rule (\ref{eq:cp}), we can therefore determine
the form factors $F^r(K^2)$ if we can measure the $K^2$-dependence of the fully
differential cross-section $d^3\D\s / dQ^2 dx_e dK^2$.

We now discuss whether such measurements are feasible at present and future
$e^+ e^-$ colliders. For this purpose, we consider two accelerators in detail -- the
International Linear Collider (ILC) and the proposed high-luminosity $B$ factory
SuperKEKB. The contrasting machine parameters illustrate clearly the main issues
involved in measuring the first moment sum rule.

At the time the sum rule was proposed in ref.\cite{NSVone}, the luminosity
available from the then current accelerators was inadequate for measuring
the sum rule. For example, for a polarised version of LEP operating at $s=10^4 {\rm
GeV}^2$ with an annual integrated luminosity of $L=100{\rm pb}^{-1}$, and
choosing the cut at $Q_{\rm min}^2 = 10{\rm GeV}^2$, we find $\s \simeq 35{\rm pb}$
and $\D\s/\s \simeq 0.01$. However, the corresponding annual event rate would be
$3.5\times 10^3$ and the statistical significance only $\sqrt{L\s} \D\s/\s
\simeq 0.5$, so even a reliable measurement of the spin asymmetry could not be
made. 

Clearly, a hugely increased luminosity is required and this has now become
available with proposals for machines with projected annual integrated luminosities
measured in inverse attobarns. However, as noted in ref.\cite{NSVone}, if this
increased luminosity is associated with increased CM energy, then the $1/s$ factor
in the spin asymmetry (\ref{eq:eb}) sharply reduces the possibility of extracting
a measurement of $g_1^\c$. For this reason, it was already recognised in 
ref.\cite{NSVone} that the best future colliders for studying the sum rule would
be high-luminosity $B$ factories.

\FIGURE
{\epsfxsize=6.5cm\epsfbox{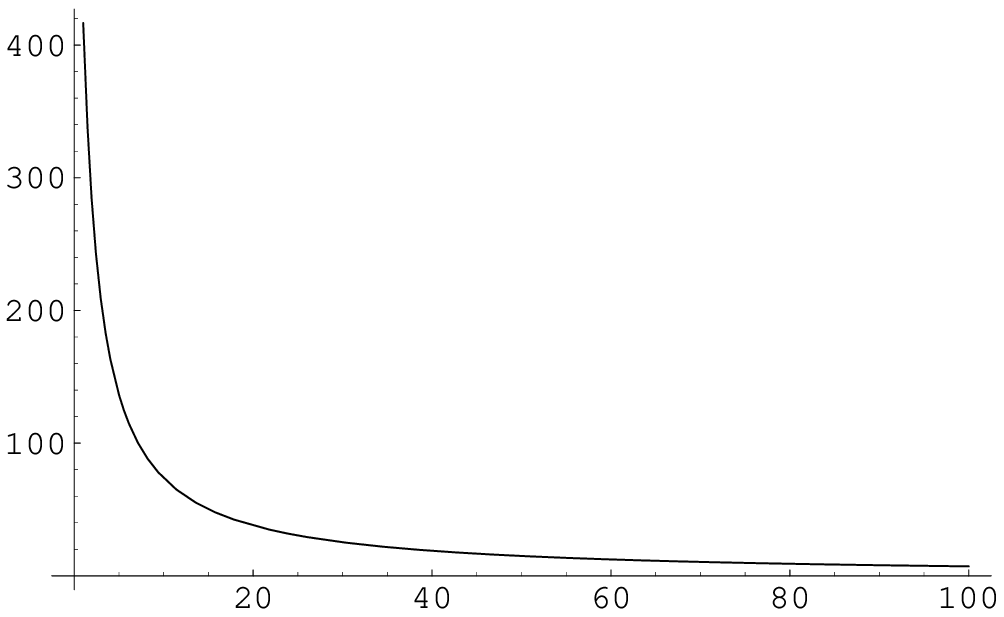} \hskip1cm
\epsfxsize=6.5cm\epsfbox{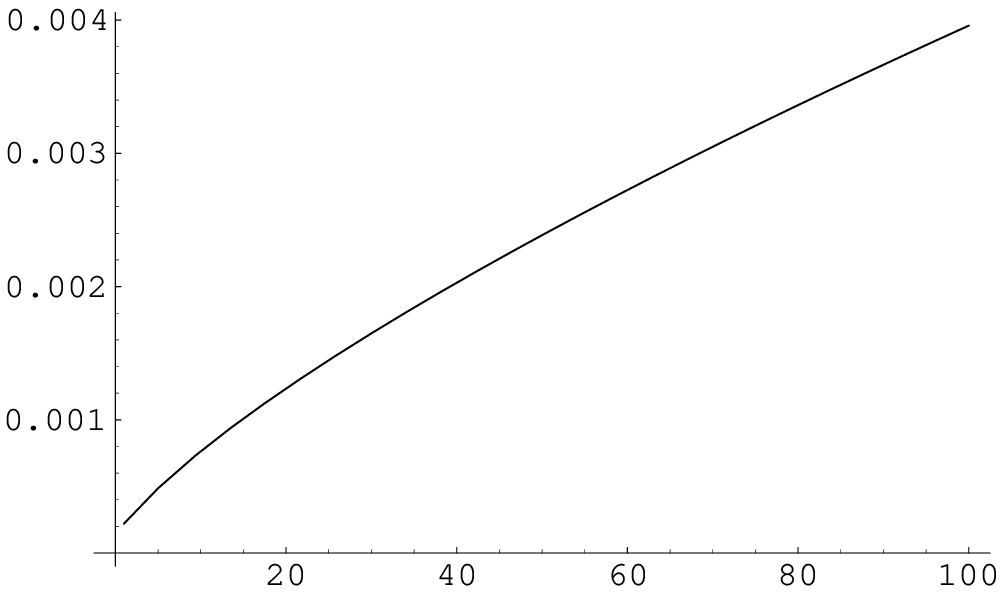}
\caption{The first graph shows the fall in total cross-section $\s$ (in pb) at the 
ILC as the experimental cut $Q_{\rm min}^2$ is varied from 1 to 100${\rm GeV}^2$. 
The second graph shows the spin asymmetry $\D\s/\s$ rising over the same range of 
$Q_{\rm min}^2$.}
\label{Fig:2}} 

\FIGURE
{\epsfxsize=6.5cm\epsfbox{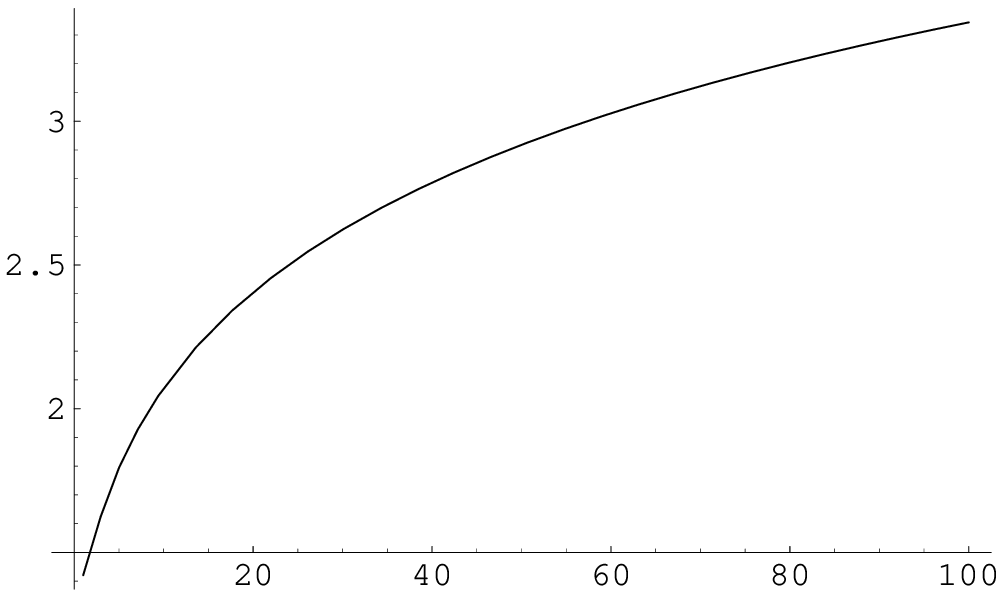} 
\caption{Plot of the statistical measure $\sqrt{L\s} \D\s/\s$ for  $Q_{\rm min}^2$
between 1 and 100${\rm GeV}^2$ at the ILC.}
\label{Fig:3}} 
Now consider the accelerator parameters of the ILC. This will operate initially at 
a CM energy of $500{\rm GeV}$  ($s = 2.5\times 10^5{\rm GeV}^2$) with a projected 
luminosity of $10^{34}{\rm cm}^{-2} s^{-1}$, corresponding to an annual integrated 
luminosity of $0.1 {\rm ab}^{-1}$ \cite{ILCone,ILCtwo}.
For simplicity, we assume here that this luminosity could
be achieved with the ILC running with polarised beams. We now analyse how
to optimise a measurement of $g_1^\c$ by varying the experimental cuts, in
particular $Q_{\rm min}^2$. In Fig.~2, we have plotted the cross-section
$\s$ (in pb) and spin asymmetry $\D\s/\s$ for values of $Q_{\rm min}^2$ varying
from 1 to $100{\rm GeV}^2$. The sharp fall-off in the cross-section is determined by 
the $1/Q_{\rm min}^2$ factor in eq.(\ref{eq:ed}). On the other hand, in this range, 
the spin asymmetry rises with increasing $Q_{\rm min}^2$ because of the 
corresponding factor in eq.(\ref{eq:ee}). The absolute value of $\D\s/\s$ is
kept relatively small because of the $1/s$ dependence of the spin asymmetry on the
CM energy. Optimising the cut on $Q_{\rm min}^2$ is therefore a balance between
keeping the total event rate high and maximising the spin asymmetry. If we plot
$\sqrt{L\s} \D\s/\s$ for this same range of $Q_{\rm min}^2$ (Fig.~3), we
see that it rises monotonically -- in fact it reaches a maximum only at 
$Q_{\rm min}^2 \sim 10^3{\rm GeV}^2$ where the cross-section has fallen
to a mere 0.5pb. 

As a reasonable compromise between event rate and spin asymmetry,
we could choose to take the cut at $Q_{\rm min}^2 \simeq 50{GeV}^2$. 
This corresponds to $\s \simeq 15{\rm pb}$ and $\D\s/\s \simeq 0.002$. The
annual event rate is $1.5\times 10^6$ with $\sqrt{L\s} \D\s/\s \simeq 3$,
which would allow a measurement of the spin asymmetry itself. However, as 
we see from eq.(\ref{eq:bm}), this determines only the $n=0$ moment
of $g_1^\c$, integrated over $K^2$. A detailed study of the first moment sum 
rule itself would require a much greater $\D\s/\s$.

This leads us to consider instead the new generation of ultra-high luminosity
$e^+ e^-$ colliders. Although these are envisaged as $B$ factories, these colliders 
operating with polarised beams would, as we now show, be extremely valuable 
for studying polarisation phenomena in QCD. As an example of this class, we take
the proposed SuperKEKB collider. (The analysis for PEPII is very similar,
the main difference being the additional ten-fold increase in luminosity
in the current SuperKEKB proposals.)

\FIGURE
{\epsfxsize=6.5cm\epsfbox{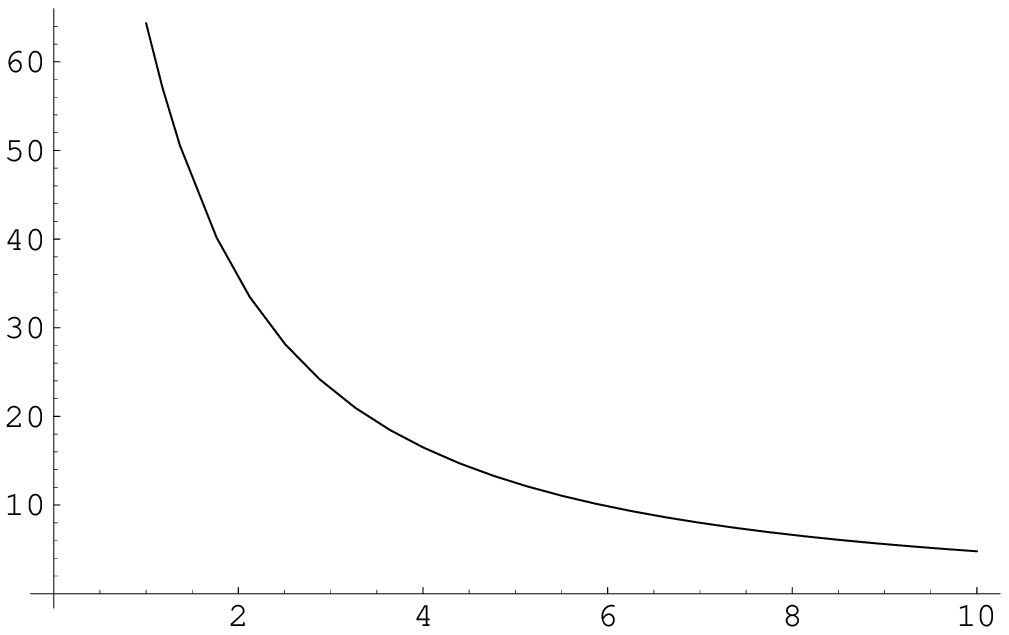} \hskip1cm
\epsfxsize=6.5cm\epsfbox{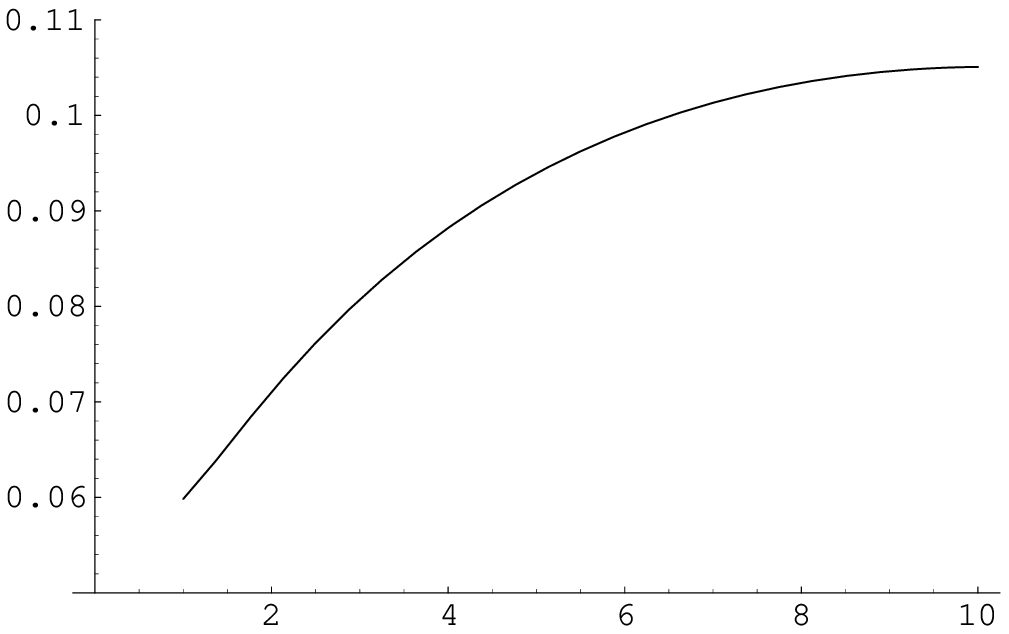}
\caption{The first graph shows the total cross-section $\s$ (in pb) at SuperKEKB 
as the experimental cut $Q_{\rm min}^2$ is varied from 1 to 10${\rm GeV}^2$. 
The second graph shows the spin asymmetry $\D\s/\s$ over the same range of 
$Q_{\rm min}^2$.}
\label{Fig:4}} 

\FIGURE
{\epsfxsize=6.5cm\epsfbox{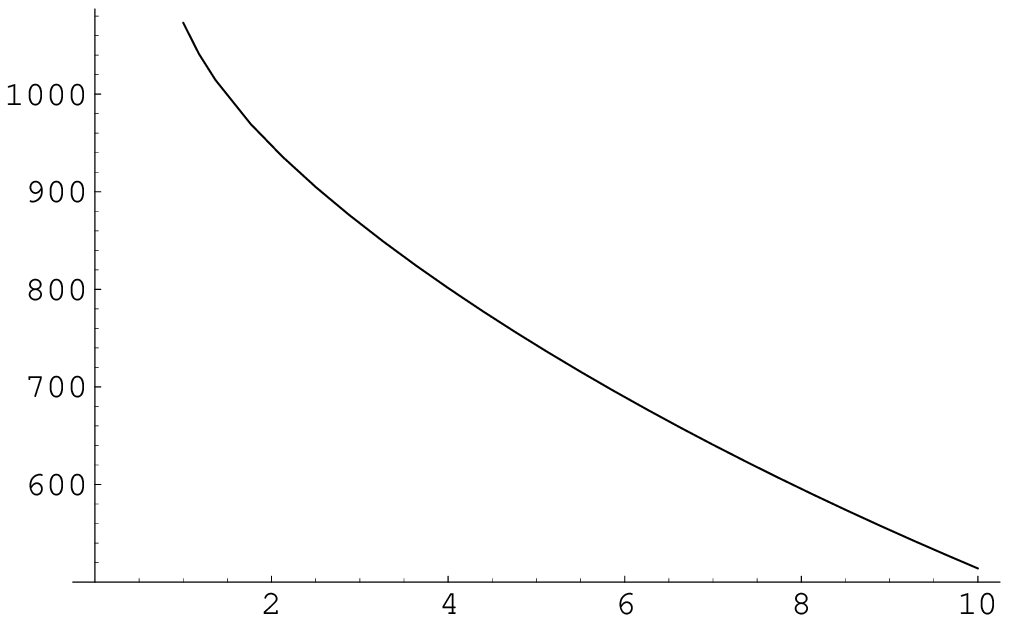} 
\caption{Plot of the statistical measure $\sqrt{L\s} \D\s/\s$ for  $Q_{\rm min}^2$
between 1 and 10${\rm GeV}^2$ at SuperKEKB.}
\label{Fig:5}} 
SuperKEKB is an asymmetric $e^+ e^-$ collider with $s= 132{\rm GeV}^2$,
corresponding to electron and positron beams of 8 and 3.5GeV respectively.
The design luminosity is $5\times 10^{35}$ ${\rm cm}^{-2} s^{-1}$, which
gives an annual integrated luminosity of $5{\rm ab}^{-1}$ \cite{SuperKEKB}. 
To see the effects of
the experimental cut on $Q_{\rm min}^2$ in this case, we again plot the total
cross-section and the spin asymmetry in Fig.~4, this time for the
range of $Q_{\rm min}^2$ from 1 to $10 {\rm GeV}^2$. As before, in this range
$\s$ is falling like $1/Q_{\rm min}^2$ while $\D\s/\s$ rises to what is
actually a maximum at $1/Q_{\rm min}^2=10{\rm GeV}^2$. On the other hand, the
statistical significance $\sqrt{L\s} \D\s/\s$ falls monotonically, though is 
orders of magnitude improved on the corresponding plot for even the ILC (Fig.~5).

Taking $Q_{\rm min}^2 =5{\rm GeV}^2$, we find $\s \simeq 12.5{\rm pb}$
with spin asymmetry $\D\s/\s \simeq 0.1$. The annual event rate
is therefore $6.25 \times 10^7$, with $\sqrt{L\s} \D\s/\s \simeq 750$.
This combination of a very high event rate and the large $10\%$ spin asymmetry
means that SuperKEKB has the potential not only to measure $\D\s$ but to
access the full first moment sum rule for $g_1^\c$ itself. Recall from
eq.(\ref{eq:ef}) that to measure $\int_0^1 dx~g_1^\c(x,Q^2;K^2)$ we need
not just $\D\s$ but the fully differential cross-section w.r.t.~not only
$x_e$ and $Q^2$, but also $K^2$ if the interesting non-perturbative QCD physics
is to be accessed. To measure this, we need to divide the data into sufficiently
fine $K^2$ bins in order to plot the explicit $K^2$ dependence of $g_1^\c$,
while still maintaining the statistical significance of the asymmetry.
The ultra-high luminosity of SuperKEKB ensures that the event rate is
sufficient, while its moderate CM energy means that the crucial spin asymmetry
is not overly suppressed by its $1/s$ dependence.

Our conclusion is that the new generation of ultra-high luminosity, moderate 
energy $e^+ e^-$ colliders, currently conceived as $B$ factories, could also
be uniquely sensitive to important QCD physics if run with polarised beams. 
In particular, they appear to be the only accelerators capable of accessing
the full physics content of the sum rule for the first moment of the
polarised structure function $g_1^\c(x,Q^2;K^2)$. The richness of this physics,
in particular the realisation of chiral symmetry breaking, the manifestations
of the axial $U_A(1)$ anomaly and the role of non-perturbative gluon dynamics,
provides a strong motivation for giving serious consideration to an attempt to
measure the $g_1^\c$ sum rule at these new colliders.

\acknowledgments

I would like to thank S.~Narison and G.~Veneziano for their collaboration on
ref.\cite{NSVone} and their helpful comments on this paper.

\end{document}